\documentclass{elsart}
\usepackage{graphicx}
\usepackage{amssymb}
\newcommand {\ignorethis}[1]{}
\newcommand {\myskip} {\hspace{.15em}}
\newcommand {\given} {\; | \myskip}
\newcommand {\beau} {\begin{eqnarray*}}
\newcommand {\eeau} {\end{eqnarray*}}

\newcommand{\dlik} {d^{\,lik}} %% targeted based on LR
\newcommand{\ddif} {d^{\,dif}} %% targeted based on dif of score averages

\begin{document}
\begin{frontmatter}


\title{Targeted Event Detection}


\author[STATaddress]{Werner Stuetzle}
\ead{wxs@uw.edu}
\ead[url]{http://www.stat.washington.edu/wxs}
\author[APLaddress,STATaddress]{Donald B.~Percival\corauthref{cor1}}
\ead{dbp@apl.washington.edu}
\ead[url]{http://faculty.washington.edu/dbp}
\corauth[cor1]{Corresponding Author}
\author[STATaddress,APLaddress]{Caren Marzban}
\ead{marzban@stat.washington.edu}
\ead[url]{http://faculty.washington.edu/marzban}
% 
%\author[STATaddress]{Albert Y.~Kim}
%\ead{albert@stat.washington.edu}
%\ead[url]{http://www.stat.washington.edu/albert}
%
\address[STATaddress]{Department of Statistics,
Box 354322,\\
University of Washington,
Seattle, WA 98195--4322, USA
}
\address[APLaddress]{Applied Physics Laboratory,
Box 355640,\\
University of Washington,
Seattle, WA 98195--5640, USA
}

\begin{abstract} 

We consider the problem of event detection based upon a (typically
multivariate) data stream characterizing some system.  Most of the
time the system is quiescent -- nothing of interest is happening --
but occasionally events of interest occur. The goal of event detection is
to raise an alarm as soon as possible after the onset of an
event. A simple way of addressing the event detection problem is to look
for changes in the data stream and equate ``change'' with ``onset of
event''. However, there might be many kinds of changes in the stream
that are uninteresting. We assume that we are given a segment of the
stream where interesting events have been marked. We propose a method
for using these training data to construct a ``targeted'' detector
that is specifically sensitive to changes signaling the onset of
interesting events.

\end{abstract}

\end{frontmatter}

% main text
%%%%%%%%%%%%%%%%%%%%%%%%%%%%%%%%%%%%%%%%%%%%%%%%%%%%%%%%%%%%%%%%%
%%%%%%%%%%%%%%%%%%%%%%%%%%%%%%%%%%%%%%%%%%%%%%%%%%%%%%%%%%%%%%%%%
\section{Introduction}
\label{sec:intro}

We consider the problem of event detection
based upon a (typically multivariate) data stream characterizing some system.
Examples include sensor readings for a patient in an intensive care unit, video images
of a scene, and sales records of pharmacies. Most of the time the system
is quiescent -- nothing of interest is happening -- but occasionally
events of interest occur: a patient goes into shock, an intruder
appears, or pharmacies in some geographic area experience increased
demand for some medications. The goal of event detection is to raise an
alarm as soon as possible after the onset of an event.

A simple way of addressing the event detection problem is to look for
changes in the data stream and equate ``change'' with ``onset of
event''.  The assumption is that, once an alarm rings, a human will
enter the loop and decide whether an event of interest did in fact
occur. If not, then the system issued a false alarm. If an event is in
progress, then the human will monitor the system till the event
ends.
Under this assumption the second alarm caused by the change from
``event'' to ``quiescent period'' would not count as a false alarm.

Changes in the data stream can be detected by comparing the
distribution of the most recent observations (the current set) with
the distribution of previous observations (the reference set). Let $T$
denote the current time. A simple approach is to choose window sizes
$C$ and $R$, and use a two-sample test $S$ to compare the observations
in the current set ${\mathcal C}_T = \{x_{T-C+1},\ldots, x_T\}$ with
the observations in the reference set ${\mathcal R}_T =
\{x_{T-C-R+1},\ldots,x_{T-C}\}$. When the test statistic $S({\mathcal
  R}_T, {\mathcal C}_T)$ exceeds a chosen threshold $\tau$, we ring
the alarm. The threshold controls the tradeoff between false alarms
and missed detections. Abstracting away details, a change
detector can be defined as a combination of a detection algorithm
mapping the multivariate input stream $x_T$ into a univariate
{\it detection stream\/} $d_T$, and an alarm threshold $\tau$. The only
restriction is that $d_T$ can depend only on input observed up to time
$T$.

A weakness of the approach to event detection outlined above is the
equating of ``onset of event'' with ``change'': there might be many
kinds of changes in the stream that do not signal the onset of an
event of interest.  If we detect changes by running two-sample tests,
the weakness can be expressed in terms of the power characteristics of
the test $S$.  We want $S$ to have high power for discriminating
between data observed during quiescent periods and data observed
at the onset of an interesting event, and low power against all other alternatives. The
difficulty is that it can be hard to ``manually'' design such a test,
especially in a multivariate setting.

In a previous paper~\cite{Kim09a} we argued that realistically
assessing the performance of a change detector and choosing the
threshold $\tau$ for a desired false alarm rate requires labeled data.
By this we mean a segment $x_1, \ldots, x_n$ of the data stream with
labels $y_1, \ldots, y_n$, where $y_i=1$ if $x_i$ is observed during
an event and $y_i=0$ if $x_i$ is observed during a quiescent period.
The assumption that we have labeled training data begs a question:
shouldn't we use these data for designing rather than merely
evaluating a detector?  In this paper we propose a way of injecting
labeled data into the design phase of an event detector. We refer to
this process as training or ``targeting'' the detector.

The remainder of this paper is organized as follows: In
Section~\ref{sec:event-detection} we describe the basic idea behind
targeted event detection and contrast it with untargeted event
detection. Targeting converts the problem of detecting a change in the
data stream signaling the onset of an event to the problem of detecting
a positive level shift in a univariate stream; we address this problem
in Section~\ref{sec:level-shift}. In Section~\ref{sec:evaluation} we
briefly sketch an adaptation of ROC curves to event detection proposed
in~\cite{Kim09a}. In Section~\ref{sec:illustrationUnivariate} we illustrate the
effect of targeting in a simple situation where the data stream is
univariate and the observations are independent. A more realistic
multivariate example is presented in
Section~\ref{sec:illustrationImage}. Section~\ref{sec:discussion} concludes the
paper with a summary and discussion.

%%%%%%%%%%%%%%%%%%%%%%%%%%%%%%%%%%%%%%%%%%%%%%%%%%%%%%%%%%%%%%%%%

\section{Targeting an Event Detector}
\label{sec:event-detection}

We assume we are given a segment $x_1, \ldots, x_n$ of a (possibly
multivariate) data stream together with class labels $y_1, \ldots,
y_n$, where $y=1$ if $x_i$ was observed during an event of interest,
and $y_i=0$ otherwise. We use these training data to target the event
detector.

The key step in our targeting method is to train a classifier on the
labeled data. The classifier produces a classification score $s_i$ for
each $x_i$, with large values indicating $y_i=1$; i.e., $x_i$ was
observed during an event.

By construction, onset of an event is signaled by a positive shift in
the score stream. We are now left with the simpler problem of
detecting a positive level shift in a univariate stream; two
univariate change detectors mapping scores into a detection stream
$d_T$ are described in Section~\ref{sec:level-shift}. We raise an
alarm whenever the detection stream produced by the univariate
detection algorithm exceeds a chosen threshold $\tau$. The choice of
$\tau$ controls the tradeoff between false alarms and missed
events. Note that labeled data are needed only for the training phase
and not during the operation of the change detector.

\begin{figure}[t]
\centerline{
\hspace*{-1.05in} \includegraphics[height=11.0in, width=8in]{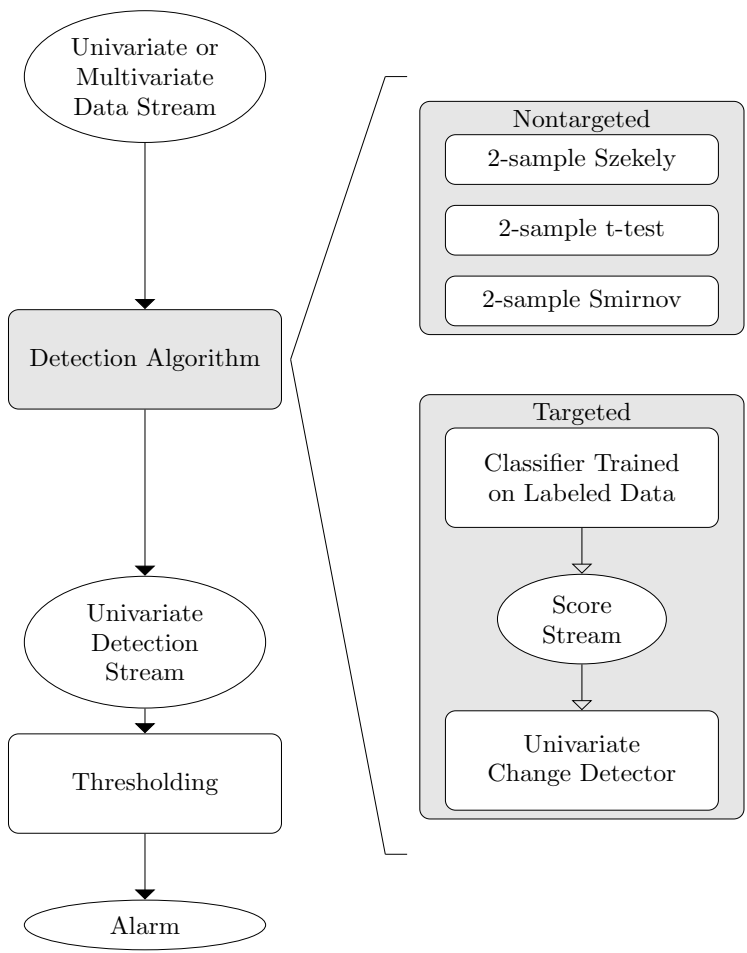}
} 
\vspace{-4.0in} 
\caption{Flow chart showing the general structure of a change detector
  (left) and two versions of detection algorithms --- targeted and
  untargeted (right).}
\label{fig:flowchart}
\end{figure}

It is useful to contrast targeted and untargeted event
detection.
Figure~\ref{fig:flowchart} shows a flowchart
contrasting the two approaches.
In targeted event detection, the detection
algorithm transforming the data stream into a univariate detection
stream is based on a scoring procedure derived from previously
observed labeled training data. In untargeted event detection it is
up to the designer of the detector to choose a two-sample test
sensitive against changes signaling the onset of an event. The
standard choices like the multivariate $T$-test and the $F$-test have
power only against location and scale changes, respectively, whereas
the change in the data stream signaling the onset of events might be of
a more complex nature. There are omnibus two sample tests, like
Szekely's test~\cite{Asla05a,Rubi05a,Szek89a,Szek00a,Szek04a}, that
are consistent against all alternatives; however, their power
characteristics might not be well matched to the problem at hand.

%%%%%%%%%%%%%%%%%%%%%%%%%%%%%%%%%%%%%%%%%%%%%%%%%%%%%%%%%%%%%%%%%

\section{Detecting a Level Shift in the Score Stream}
\label{sec:level-shift}

Targeting transforms the problem of detecting a change in a (typically
multivariate) data stream signalling the onset of an event into the
simpler problem of detecting a positive level shift in the univariate
score stream generated by the classifier.
An obvious approach is to
compare the average scores in the current and reference windows,
leading to the detection stream
\begin{equation}\label{eq:FirstDetectStream}
\ddif_T
=
\frac{1}{C}\sum_{x_i\in{\mathcal C}_T} s_i - 
\frac{1}{R}\sum_{x_i\in{\mathcal R}_T} s_i\,.
\end{equation}

An alternative approach is motivated by likelihood ratio tests.
Suppose, for the moment, that observations in the data stream were
independent and that we knew the class conditional densities
$p_0(x) = p(x\vert y=0)$ and $p_1(x)$.
The likelihood ratio statistic for
testing the null hypothesis that all of the observations in ${\mathcal C}_T$
and ${\mathcal R}_T$ come from $p_0$ against the alternative hypothesis
that all of the observations in ${\mathcal R}_T$ come from $p_0$ and all
of those in ${\mathcal C}_T$ come from $p_1$ is
\[
\lambda
=
\frac{\prod_{x_i\in{\mathcal R}_T} p_0(x_i)\prod_{x_i\in{\mathcal C}_T} p_1(x_i)}
{\prod_{x_i\in{\mathcal R}_T\cup {\mathcal C}_T} p_0(x_i)}
=
\prod_{x_i\in{\mathcal C}_T}
\frac{ p_1(x_i)} { p_0(x_i)}\,.
\]
We reject the null hypothesis for large values of $\lambda$.  The log
likelihood ratio can be written as a function of $p(y=1\vert x)$:
\[
\log \lambda
=
\sum_{x_i\in{{\mathcal C}_T}}
\log\left(
\frac{ p_1(x_i)}{ p_0(x_i)}
\right)
=
\sum_{x_i\in{{\mathcal C}_T}}
\left[
\log\left(
\frac{ p(y=1\vert x_i)}{ p(y=0\vert x_i)}
\right) +
\log\left(
\frac{p(y=0)}{p(y=1)}
\right)
\right].
\]
Regarding $s_i$ as an estimate for 
to be $p(y=1\vert x_i)$, this argument motivates
the detection stream
\begin{equation}\label{eq:SecondDetectStream}
\dlik_T
=
\sum_{x_i\in{{\mathcal C}_T}}
\log\left(
\frac{s_i}{1 - s_i}
\right),
\end{equation}
which is independent of the reference set.
(We can drop the term involving $p(y=0)/p(y=1)$
since it does not depend on the data stream.)

%%%%%%%%%%%%%%%%%%%%%%%%%%%%%%%%%%%%%%%%%%%%%%%%%%%%%%%%%%%%%%%%%
\section{Evaluation of Event Detectors}
\label{sec:evaluation}

An event detector can make two kinds of errors: it can issue false
alarms, or it can signal events with undue delay or not at
all. Raising an alarm soon after the start of an event is crucial for
event detection: if the alarm occurs too long after the start, the horse
will have left the barn, and the alarm is useless.  Also, changes within
events or transitions from events to quiescent periods are not of
interest. Following Kim~\etal~\cite{Kim09a}, we define an event to be
successfully detected if the detection stream exceeds the alarm
threshold $\tau$ at least once within a tolerance window of size $W$
after the onset of the event. We define the hit rate $h(\tau)$ as the
proportion of successfully detected events.  The false alarm rate
$f(\tau)$ is simply the proportion of times in the quiescent periods
during which the detection stream exceeds the alarm threshold.  There
is no penalty for raising multiple alarms during an event. Our
definitions for $h(\tau)$ and $f(\tau)$ are admittedly simple, and
others might be better in scenarios not involving event detection.

We can summarize the performance of a change detection algorithm by
plotting the hit rate $h(\tau)$ versus the false alarm rate $f(\tau)$
as we increase the alarm threshold $\tau$. Both $h(\tau)$ and
$f(\tau)$ are monotonically non-increasing functions of $\tau$. The
graph of the curve $\tau \longrightarrow (f(\tau), h(\tau))$ is a
monotonically non-decreasing function of $f(\tau)$. We call this curve
the ROC curve for the algorithm since it is similar to the standard
ROC curve used to evaluate binary classifiers~\cite{Fawc06a}.

It is useful to compare the performance of a detection algorithm with
the performance of the proverbial monkey who ignores the data and signals an alarm
with probability $\alpha\in[0,1]$ independently at each time $T$.
Clearly the false alarm rate for the monkey is $\alpha$.  The rate at
which the monkey will successfully flag an event is given by the
probability that an alarm is raised at least once within the tolerance
window of size $W$, which is governed by a binomial distribution with
parameters $W$ and $\alpha$. The ROC curve of the monkey is thus
$\alpha \longrightarrow (\alpha, 1- (1 -\alpha)^W)$.

%%%%%%%%%%%%%%%%%%%%%%%%%%%%%%%%%%%%%%%%%%%%%%%%%%%%%%%%%%%%%%%%%

\section{Illustration: Targeted Event Detection in a Univariate Stream}
\label{sec:illustrationUnivariate}

To illustrate the benefits of targeted event detection we consider a
simple simulated example where the data stream consists of independent
univariate observations. The density $p_0$ of observations during
quiescent periods is taken to be standard Gaussian. The density $p_1$
of observations during events is taken to be a mixture of two
symmetric components designed to also have zero mean and unit variance:
\[
p_1(x) = \frac{1}{2\sigma}
\left[
p_0\left(\frac{x-\mu}{\sigma}\right) + p_0\left(\frac{x+\mu}{\sigma}\right)
\right],
\]
with $\mu = 0.9$ and $\sigma^2 = 0.19$ (see Figure~\ref{fig:twoPDFs}).
Standard two-sample tests for changes in location or scale will have
poor power here since $p_0$ and $p_1$ have the same mean and
variance.

Given sufficient training data labeled as coming from events ($p_1$)
and quiescent periods ($p_0$), we can estimate both $p_0$ and $p_1$ to
any desired degree of accuracy.  Assuming for simplicity that both
densities are known perfectly, we can take the score stream to be
\[
s_i
= p(y=1\vert x_i) 
= 
\frac{p(y=1,x_i)}{p(x_i)}
=
\frac{\pi_1 p_1(x_i)}{(1-\pi_1) p_0(x_i) + \pi_1 p_1(x_i)}\,,
\]
where $\pi_1 = p(y = 1)$.

The detection stream defined in Equation~(\ref{eq:SecondDetectStream}) then
becomes
\[
\dlik_T
=
\sum_{x_i\in{{\mathcal C}_T}}
\log\left(
\frac{s_i}{1 - s_i}
\right)
=
C
\log\left(
\frac{\pi_1}{1-\pi_1}
\right)
+
\sum_{x_i\in{{\mathcal C}_T}}
\log\left(
\frac{p_1(x_i)}{p_0(x_i)}
\right).
\]
Because changing $\pi_1$ results in a level shift of $\dlik_T$, the
graph of the ROC curve for the detector does not depend on $\pi_1$.
Setting $\pi_1=1/2$ for convenience yields
\[
\dlik_T
=
\sum_{x_i\in{{\mathcal C}_T}}
\log\left(
\frac{p_1(x_i)}{p_0(x_i)}
\right),
\]
which can be interpreted as a log likelihood ratio test statistic.
Figure~\ref{fig:threeStreams} shows examples of the streams
$x_i$, $s_i$ and $\dlik_T$ with $C=20$.

In the following we assume for simplicity that the length of events is
large relative to the size $W$ of the tolerance window, and the
spacing between events is large relative to the combined size $C+R$
of the current and reference windows.  For given $W$ and alarm
threshold $\tau$ we can estimate the false alarm rate $f(\tau)$ and
the hit rate $h(\tau)$ associated with $\dlik_T$ using Monte Carlo experiments.
We estimate $f(\tau)$ by computing $\dlik_T$ for a stream of data
drawn exclusively from $p_0$ and by determining the proportion of time
that $\dlik_T$ exceeds $\tau$.
To determine the hit rate, suppose that an event starts at time
$T$ and has a duration at least as long as the tolerance window; i.e.,
$x_i$ for $i=T, T+1, \ldots, T+W-1$ are drawn from $p_1$.  Suppose
also that $x_i$ for $i=T-C,\ldots, T-1$ are drawn from $p_0$.  Since
we declare an event to be successfully detected if $\dlik_T$ exceeds
$\tau$ at least once within the tolerance window, we can estimate the
hit rate by repetition of the following four steps:
% (1) sample $x_{T-C+1},\ldots, x_{T-1}$ from $p_0$; \\
% Isn't there a fence post error
\begin{enumerate}
\item
sample $x_{T-C},\ldots, x_{T-1}$ from $p_0$;
\item
sample $x_T, x_{T+1}, \ldots, x_{T+W-1}$ from $p_1$;
\item
form the detection stream $\dlik_{T}, \dlik_{T+1}, \ldots \dlik_{T+W-1}$; and
\item
see if any of these $W$ values exceed the threshold $\tau$.
\end{enumerate}
The solid black curve in Figure~\ref{fig:illust-rocs} is the ROC curve for
the detection stream $\dlik_T$ with $W=C=20$.

To illustrate the benefit of targeting we also consider a detection
stream based on a two-sample Smirnov test~\cite{Cono99a}.  This
nonparametric test is designed to test for distributional differences
between two independent random samples, which in our case would
consist of the observations in the current set ${\mathcal C}_T$ and
reference set ${\mathcal R}_T$. Since $\dlik_T$ depends only on
${\mathcal C}_T$, it is convenient to remove the dependence of the
Smirnov test on ${\mathcal R}_T$ by presuming that $R$ is sufficiently
large so that $p_0$ is known to arbitrary precision. This allows us to
replace the two-sample Smirnov test with a one-sample Kolmogorov
goodness-of-fit test against the null hypothesis
$p_0$~\cite{Cono99a}. The solid gray curve in Figure~\ref{fig:illust-rocs} is
the ROC curve of the untargeted detector based on the Kolmogorov
test. The dashed curve is the ROC curve for the monkey ignoring the
data and signaling an alarm with probability $\alpha\in[0,1]$
independently at the each time $t$. 

We see that the targeted detector performs much better than the monkey
and the untargeted detector. The untargeted detector performs only
marginally better than the monkey for very small false alarm rates and
is actually {\it worse\/} for moderate false alarm rates! This might
seem surprising --- after all the Kolmogorov test does have some power
to distinguish $p_0$ from $p_1$. The reason is that the stream
$\dlik_T$ is correlated, while the monkey's coin tosses are not. Here
is a heuristic argument: Suppose an event starts at time $t$ and
$\dlik_t << \tau$. Because of positive auto-correlation
$\dlik_{t+1},\ldots, \dlik_{t+W}$ will likely be also less than
$\tau$, and the detector will miss the event. Now suppose on the other
hand that $\dlik_t >> \tau$. Then $\dlik_{t+1},\ldots, \dlik_{t+W}$
will likely be also greater than $\tau$, but we will not get credit
for raising the alarm multiple times. To verify that correlation
causes the poor performance of the Kolmogorov test, we can change the
procedure for estimating the hit rate: We generate new samples in
steps~(1) and~(2) each time we compute a value of the detection
stream. The resulting ROC curve (dotted) indeed is uniformly better
than the monkey.

\section{Illustration: Targeted Event Detection in an Image Stream}
\label{sec:illustrationImage}

Suppose we observe an image stream in which objects appear and move about.
Certain kinds of objects are interesting.
The presence of these objects constitutes an interesting event.
Following our basic approach to targeted event detection,
we want to construct a classifier assigning a score to each image.
This score should tend to be large
if an image shows an object of interest and small if it does not.

Image streams have several characteristics that we need to take into
account.
\begin{enumerate}
\item
They tend to be high-dimensional.
If the image resolution is $1024 \times 1024$,
we are in effect observing more that $10^6$ variables.
Even for small $100 \times 100$ images,
the dimension of the data stream is $10000$.
\item
Due to the high dimensionality,
each individual variable (pixel)
conveys relatively little information.
\item
We often do not care where an object of interest appears in the
image, and objects can move from one image to the next.
During the operational use of the event detector, objects of
interest might appear in locations where they were never seen in the
training images. Therefore the design of the event detector has to
incorporate some kind of spatial invariance.
\end{enumerate}

To accommodate these characteristics,
we assume that, during the training process,
we visually identify images showing an object of interest
and mark these objects by, e.g., placing a bounding box. The inspection
process produces a collection of boxes showing objects of interest;
we will call these ``event boxes''. Assume for simplicity that
all event boxes are of the same size, say, $m \times m$. Next we extract a
sample of $m\times m$ ``quiescent'' boxes from images taken during
quiescent periods. Using the training sample of boxes we construct a
classifier for boxes assigning a large score to event boxes and a
small score to quiescent boxes.

To decide whether an image is taken during a quiescent period or
during an even we apply the box classifier to all the boxes in the
image. If the image is $n\times n$ this results in $(n-m+1)^2$ box
scores. From these box scores we need to derive a score for the entire
image; an obvious choice is the maximum of the box scores~\cite{Kim09a}.
The problem of object detection in images has been extensively studied
in computer vision and image processing; the approach sketched above goes under
the name ``template matching''~\cite{Brun09a}.

Targeted event detection based on template matching can be very
effective. We now illustrate the approach using two simple scenarios.
In the first scenario there is one kind of interesting object and
no uninteresting objects.
In second there is one kind of interesting and one kind of uninteresting object.

\subsection{First scenario: Interesting objects only}

Consider a stream of $100 \times 100$ grey level images contaminated
by independent standard Gaussian noise. An object of interest
manifests itself by a pyramid of bright pixels with average intensity
$\mu = 3$. Figure~\ref{fig:pyramid} shows a sample image with an
interesting object. Objects can move from image to image
once they have appeared. We gather a training sample of $N_e=10$ event boxes of size
$10\times 10$ and $N_q = 10000$ quiescent boxes. In practice, event
boxes would be collected ``manually'' as described above. For our
illustration we automate this process and make it reproducible by
template matching a pyramid against $N_e$ event images and selecting
from each image the box that matches best. We then use the training
sample of event boxes and quiescent boxes to train a Fisher
discriminant rule. Assuming that events are far apart relative to the
combined size $R+W$ of the reference and tolerance windows, we can
determine the ROC curve through simulation, as in our univariate
example in Section~\ref{sec:illustrationUnivariate}.
The result for $R = 10$ and $W = C = 1$ is the
solid black line in Figure~\ref{fig:ImageROC1}
(see Section~\ref{sec:discussion} for a discussion of this choice for $W$ and $C$).
The targeted detector performs perfectly
(up to the precision imposed by the finite sample size of the simulation).
As a comparison, consider an untargeted detector
that looks for change one pixel at a time by comparing the pixel
values in the current and reference windows, and then rings an alarm
if the maximum value of the test statistic over pixels exceeds the
alarm threshold. The solid grey curve in Figure~\ref{fig:ImageROC1} shows
the ROC curve of the untargeted detector if we use absolute difference
in means as the test statistic.

This first scenario suggests that,
even in a simple situation where there are
no uninteresting objects, targeted event detection can be advantageous
because it uses information on what we are looking for. The major
advantage of targeted event detection, however, is the ability to
distinguish interesting from uninteresting events,
as the second scenario illustrates.

\subsection{Second scenario: Interesting and uninteresting objects}

To simplify analysis and understanding, we assume that at any given
time we can either see noise, or a single interesting object, or a
single uninteresting object. We call the presence of an interesting
object an interesting event, and the presence of an uninteresting
object an uninteresting event. We want to raise an alarm at the onset
of interesting events. We also assume that the durations of
events and the lengths of time between events are
both greater than or equal to $R + C$.

The probability of a false alarm at some time $T$ depends on the 
time interval $[(T - R - C + 1),\ldots T]$.
We can be in one of the following four situations.
\begin{enumerate}
\item[$A_1$:]
all images in the interval show noise;
\item[$A_2$:]
an uninteresting object is present at the beginning but not at the end
(an uninteresting event ends during the interval);
\item[$A_3$:]
an uninteresting object is present at the end but not at the beginning
(an uninteresting event starts during the interval);
\item[$A_4$:]
an uninteresting object is present during the entire interval.
\end{enumerate}
The simplifying assumptions above rule out any other patterns.

The probability of raising an alarm at time $T$ (for a given alarm
threshold) is
\[
P(F) = \sum_{i=1}^4 P(F\given A_i)\;P(A_i)\,.
\]
The conditional probabilities $P(F\given A_1)$ and $P(F\given A_4)$
are easy to obtain using simulation. Estimating the other conditional
probabilities requires a little more thought.  Consider $P(F\given
A_2)$. If an uninteresting object is visible at time 1 but not at time
$R+T$ this means an uninteresting event is ending at time 1, or at
time 2, \ldots, or at time $(R+C-1)$. Let $E_i$ stand for ``an
uninteresting event ends a time $i$''. A simple calculation shows that
\[
P(F \given A_2) = \sum P(F\given E_i)\;P(E_i \given A_2)\,.
\]
For symmetry reasons, $P(E_i \given A_2) = 1/(R+C-1)$. The conditional 
probabilities $P(F \given E_i)$ can be estimated by Monte Carlo in the
obvious way. The term $P(F\given A_3)$ is treated analogously.

The probabilities $P(A_1),\ldots,P(A_4)$ depend on the lengths of the
noise periods and of the uninteresting events. There are only two
independent parameters because for symmetry reasons $P(A_2) = P(A_3)$.
To get some intuition about the meaning of the $P(A_i)$ consider a
simple situation where noise intervals are of fixed length $N$ and
uninteresting events are of length $U$, with $N, U > R+C$. Then
\beau
P(A_1) &=& \frac{N-(R+C)+1}{N+U} \\
P(A_4) &=& \frac{U-(R+C)+1}{N+U} \\
P(A_2) &=& P(A_3) = \frac{R+C-1}{N+U}\,.
\eeau
The benefits of targeting become most apparent if uninteresting events
occur frequently. In the examples below we choose the extreme case $N
= U = R + C + 1$, leading to $P(A_1) = P(A_4) \approx 0.04$ and $P(A_2) =
P(A_3) \approx 0.46$.

Suppose that an uninteresting object manifests itself by an inverted
pyramid with average intensity $\mu = 3$. The solid black curve in
Figure~\ref{fig:ImageROC2} is the ROC curve for the targeted detector
which is close to perfect.
The solid grey curve is the ROC curve for the untargeted
detector. The targeted detector appears to be largely immune to the
occurrence of uninteresting events, while comparison of the grey
curves in Figures~\ref{fig:ImageROC1} and ~\ref{fig:ImageROC2} shows
that uninteresting events worsen the performance of the untargeted
detector. Figure~\ref{fig:ImageROC3} shows the ROC curves of the
targeted detector (solid line) and the untargeted detector (solid grey curve)
for $\mu = 5$.  Increasing the signal-to-noise ratio does not alleviate the
performance problem of the untargeted detector
except for large false alarm rates.

%%%%%%%%%%%%%%%%%%%%%%%%%%%%%%%%%%%%%%%%%%%%%%%%%%%%%%%%%%%%%%%%%

\section{Summary and Discussion}
\label{sec:discussion}

We have considered the problem of event detection based upon a (typically
multivariate) data stream characterizing some system. One of the key
challenges in automated event detection is to design an algorithm that is
sensitive to changes in the data stream signalling the onset of
interesting events but insensitive to other kinds of variability. We
have proposed a method for automating the design process. We assume
that we are given a segment of the data stream where interesting
events have been labeled. We use a (typically nonparametric)
classifier trained on the labeled data to generate a classification
rule. The classification rule maps the data stream into a
univariate score stream, where high scores indicate the occurrence
of an interesting event. We have thereby transformed the challenging problem of
detecting interesting changes in the data stream to the much simpler
problem of detecting positive level shifts in the univariate score
stream. We have illustrated our idea on a simple univariate example
with a simulated data stream and a more realistic multivariate
example. Both examples demonstrate that targeting can indeed improve
performance.

This paper suggests some avenues for future research.
For example,
the choices for the sizes $R$, $C$ and $W$
of the reference, current and tolerance windows
we made in Sections~\ref{sec:illustrationUnivariate} and~\ref{sec:illustrationImage}
were dictated mainly by the desire to illustrate
our main points as easily as possible.
The choice $W=C=1$ in Section~\ref{sec:illustrationImage}
is obviously unrealistic in practical situations,
but was convenient to assume
since it avoided the need to decide
how interesting or uninteresting objects move
from one image to the next during an event.
In general, the choice $W=C$ seems natural,
but, while it is possible to analytically demonstrate
that the ROC for $W=C$ dominates the one for $W<C$ in a simple scenario
(namely, a stream of standard Gaussian white noise
subject to a shift in its mean when an event occurs),
the choice $W>C$ is harder to rule out
(limited computer experiments suggest it might be a reasonable choice).
How to best choose $R$, $C$ and $W$ in situations
where targeted event detection is the main focus is not obvious.

Another interesting avenue for research would be
to consider the possibility of operator feedback.
Suppose, for example,
that an event detector is tuned to certain interesting events,
but, with the passage of time,
new interesting events can arise that are unlikely to raise an alarm.
Suppose also that an operator only responds to alarms raised by the event detector
and hence would be unlikely to see a false alarm
raised by new interesting events.
By raising false alarms at random times,
we can increase the probability that the operator will see new interesting events.
Assuming that there is a cost associated with responding to alarms
and a cost associated with ignoring the new interesting events,
research would be needed to determine the best strategy
for getting operator feedback that would result
in new training data for use in updating the existing event detector.

Finally, more research is needed on how best to handle
multivariate streams with high dimension,
but with less structure than image streams.
The spatial structure in images simplifies the identification of events.
The lack of a corresponding structure in other multivariate streams
can make it difficult for operators to provide feedback 
that could be used for retargeting an event detector.
Even the basic question of how to create a reasonable score stream
becomes much more difficult
when we cannot rely on preconceived notions
about the relationships between the variables in the stream.

%%%%%%%%%%%%%%%%%%%%%%%%%%%%%%%%%%%%%%%%%%%%%%%%%%%%%%%%%%%%%%%%%

\section*{Acknowledgments}
This work was funded by the U.S.~Office of Naval Research
under grant number N00014--05--1--0843.
The authors thank Albert Kim for his work on this grant.

% \appendix

% \section{}
% \label{}

%%%%%%%%%%%%%%%%%%%%%%%%%%%%%%%%%%%%%%%%%%%%%%%%%%%%%%%%%%%%%%%%%

\clearpage

%%%%%%%%%%%%%%%%%%%%%%%%%%%%%%%%%%%%%%%%%%%%%%%%%%%%%%%%%%%%%%%%%%%%%%%%%%%%

\section*{Figure Captions}
{\bf Figure~1.} {Flow chart showing the general structure of a change detector
  (left) and two versions of detection algorithms --- targeted and
  untargeted (right).
}
\par\bigskip\noindent
%%%
{\bf Figure~2.} {Standard Gaussian density $p_0$ (dashed curve) and Gaussian mixture PDF $p_1$ (solid),
also with zero mean and unit variance.
}
\par
\begin{figure}[t]
\centerline{
\hspace{0in}   \includegraphics[height=4in, width=5in]{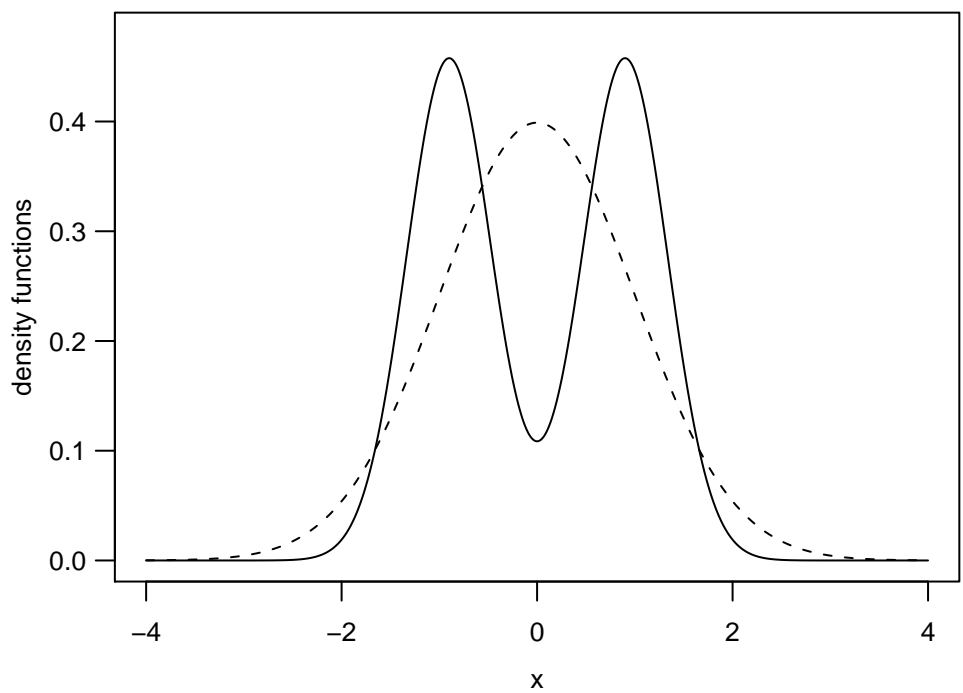}
} 
%\vspace{-2.0in} 
\caption{Standard Gaussian density $p_0$ (dashed curve) and Gaussian mixture PDF $p_1$ (solid),
also with zero mean and unit variance.}
\label{fig:twoPDFs}
\end{figure}

\begin{figure}[t]
\centerline{
\hspace{0in}   \includegraphics[height=6in, width=5in]{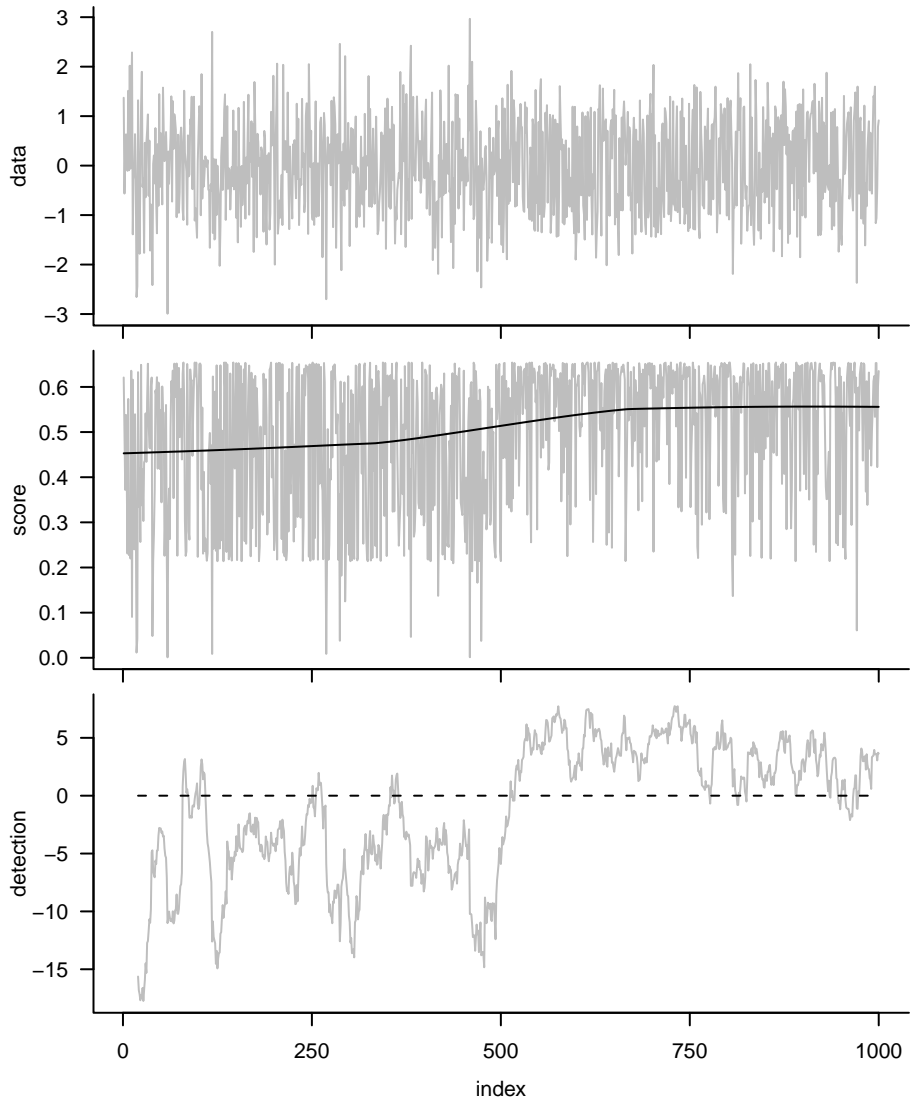}
} 
%\vspace{-2.0in} 
\caption{Data stream $x_i$ drawn from a standard Gaussian density $p_0$
for indices $i=1, \ldots, 500$
and from the Gaussian mixture $p_1$
of Fig.~\ref{fig:twoPDFs} for $i=501, \ldots, 1000$ (top);
corresponding score stream $s_i$ (middle);
and corresponding detection stream $\dlik_T$ with $C=20$ (bottom).
The black curve in the middle plot is a smooth of $s_i$
obtained by locally weighted regression~\cite{Clev79a}.
The dashed line in the bottom plot indicates the natural break between
favoring $p_0$ or $p_1$ in a log likelihood ratio test.}
\label{fig:threeStreams}
\end{figure}

\begin{figure}[t]
\centerline{
\hspace{0in}   \includegraphics[height=4in, width=5in]{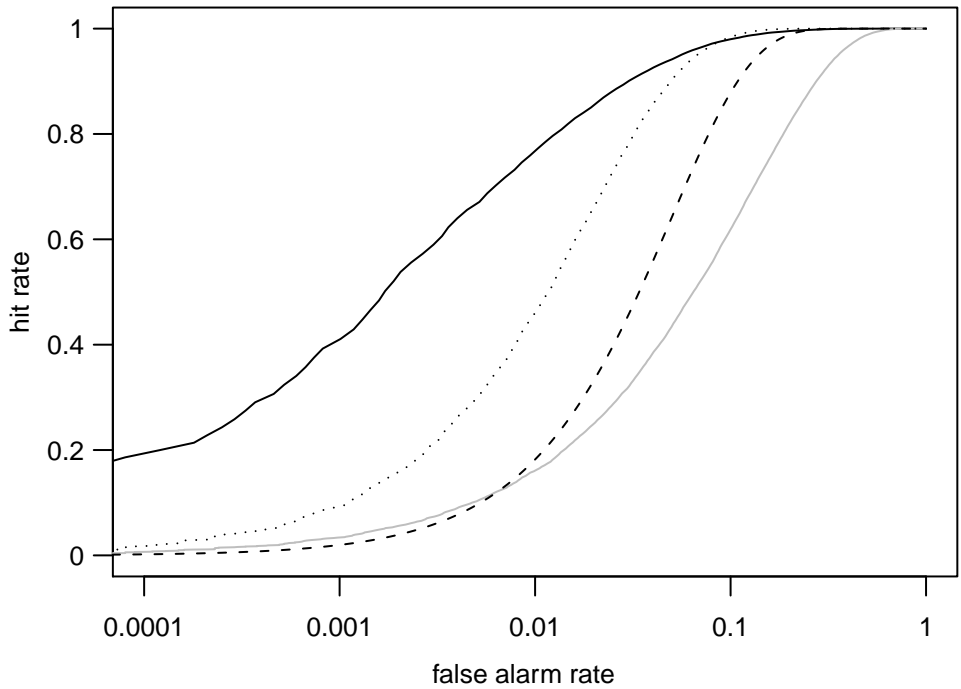}
} 
%\vspace{-2.0in} 

\caption{ROC curves for a targeted detector based on $\dlik_{T}$
  (solid dark curve), for an untargeted detector based on the
  Kolmogorov test statistic (solid gray curve), and for the monkey
  (dashed curve).  The dotted curve is the ROC curve of an
  unrealizable procedure that uses a Kolmogorov test statistic with a
  new set of independent data for each recalculation of the statistic
  within the tolerance window. The sizes of the current set $C$ and of
  the tolerance window are taken to be 20.}
\label{fig:illust-rocs}
\end{figure}

\bigskip\noindent
%%%
{\bf Figure~3.} {Data stream $x_i$ drawn from a standard Gaussian density $p_0$
for indices $i=1, \ldots, 500$
and from the Gaussian mixture $p_1$
of Fig.~\ref{fig:twoPDFs} for $i=501, \ldots, 1000$ (top);
corresponding score stream $s_i$ (middle);
and corresponding detection stream $\dlik_T$ with $C=20$ (bottom).
The black curve in the middle plot is a smooth of $s_i$
obtained by locally weighted regression~\cite{Clev79a}.
The dashed line in the bottom plot indicates the natural break between
favoring $p_0$ or $p_1$ in a log likelihood ratio test.
}
\par\bigskip\noindent
%%%
{\bf Figure~4.} {ROC curves for a targeted detector based on $\dlik_{T}$
  (solid dark curve), for an untargeted detector based on the
  Kolmogorov test statistic (solid gray curve), and for the monkey
  (dashed curve).  The dotted curve is the ROC curve of an
  unrealizable procedure that uses a Kolmogorov test statistic with a
  new set of independent data for each recalculation of the statistic
  within the tolerance window. The sizes of the current set $C$ and of
  the tolerance window are taken to be 20.
}
\par\bigskip\noindent
%%%
{\bf Figure~5.} {Grey level image of size $100\times100$ showing
uncorrelated standard Gaussian noise, to which has been
added an interesting object (the pyramid in the middle,
corrupted by noise).
}
\par
\begin{figure}[t]
\centerline{
\hspace{0in}   \includegraphics{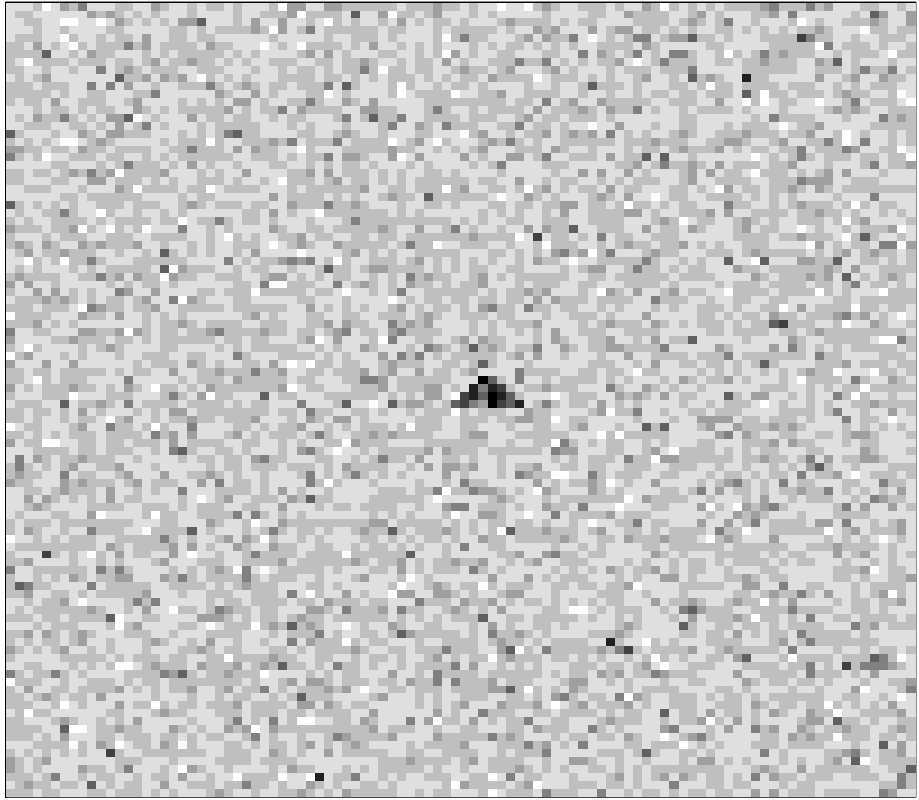}
} 
%\vspace{-2.0in} 
\caption{Grey level image of size $100\times100$ showing
uncorrelated standard Gaussian noise, to which has been
added an interesting object (the pyramid in the middle,
corrupted by noise).}
\label{fig:pyramid}
\end{figure}

\bigskip\noindent
%%%
{\bf Figure~6.} {ROC curves for targeted (solid line)
and untargeted (gray curve) detectors under the scenario 
that images contain either just Gaussian noise
or an interesting object in the presence of noise
(Fig.~\ref{fig:pyramid} is an example of the latter case).
The dashed curve is for the monkey detector.
}
\par
\begin{figure}[t]
\centerline{
\hspace{0in}   \includegraphics{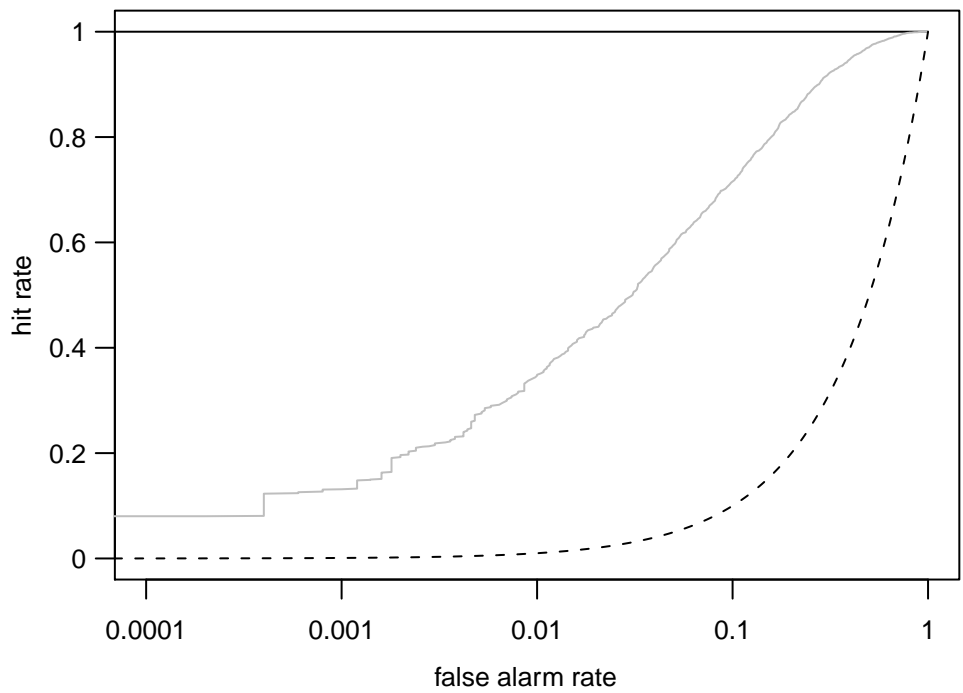}
} 
%\vspace{-2.0in} 
\caption{ROC curves for targeted (solid line)
and untargeted (gray curve) detectors under the scenario 
that images contain either just Gaussian noise
or an interesting object in the presence of noise
(Fig.~\ref{fig:pyramid} is an example of the latter case).
The dashed curve is for the monkey detector.
}
\label{fig:ImageROC1}
\end{figure}

\bigskip\noindent
%%%
{\bf Figure~7.} {As in Fig.~\ref{fig:ImageROC1},
but now under the scenario
that some of the images have an uninteresting object
(an inverted pyramid).
}
\par
\begin{figure}[t]
\centerline{
\hspace{0in}   \includegraphics{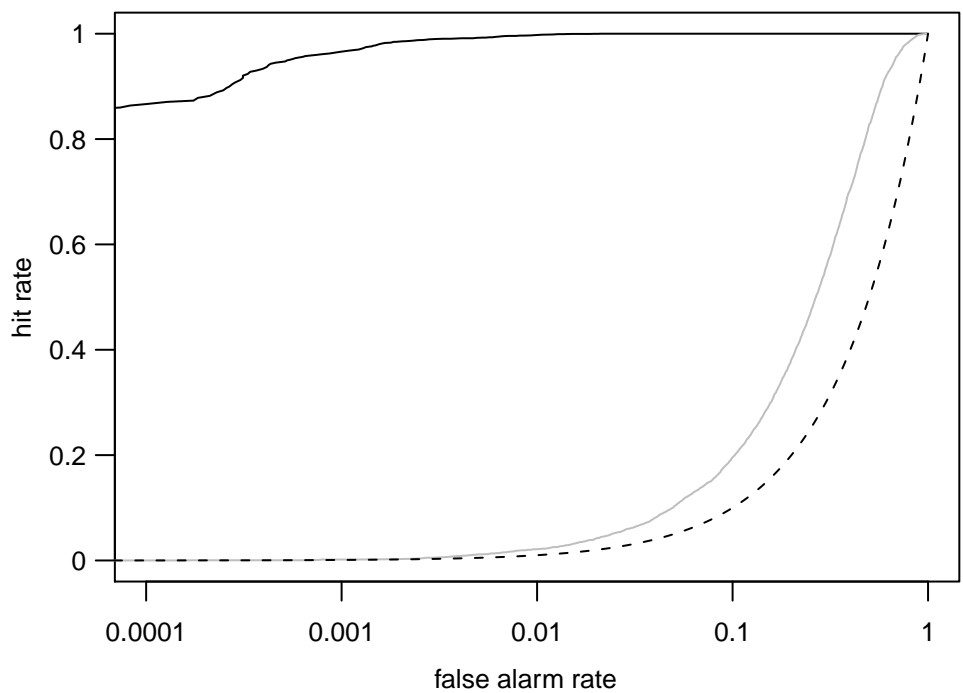}
} 
%\vspace{-2.0in} 
\caption{As in Fig.~\ref{fig:ImageROC1},
but now under the scenario
that some of the images have an uninteresting object
(an inverted pyramid).
}
\label{fig:ImageROC2}
\end{figure}

\begin{figure}[t]
\centerline{
\hspace{0in}   \includegraphics{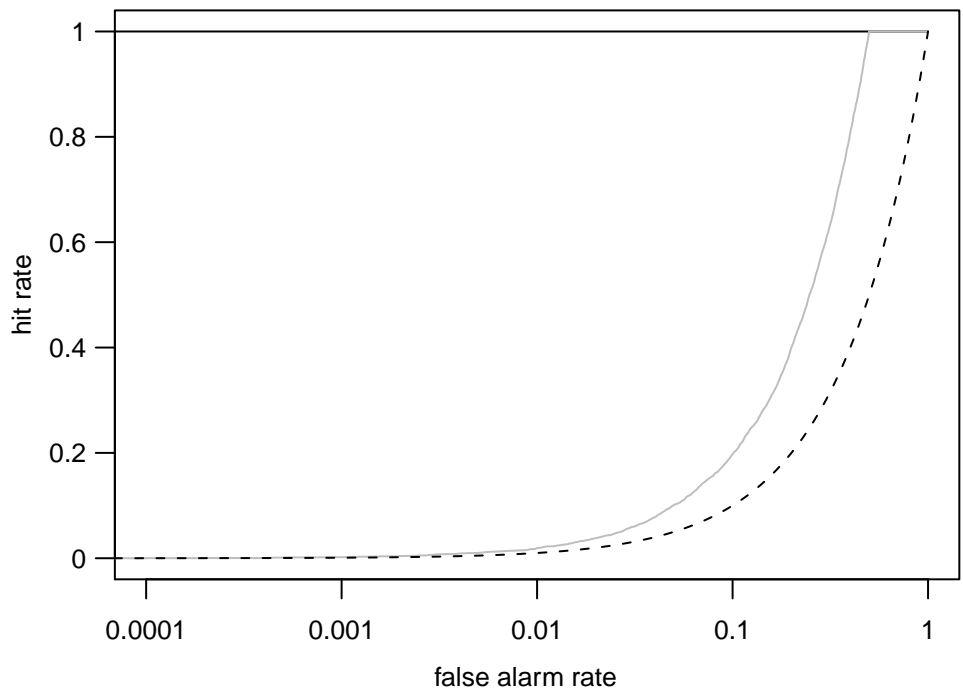}
} 
%\vspace{-2.0in} 
\caption{As in Fig.~\ref{fig:ImageROC2},
but now with a higher signal-to-noise ratio.
}
\label{fig:ImageROC3}
\end{figure}

\bigskip\noindent
%%%
{\bf Figure~8.} {As in Fig.~\ref{fig:ImageROC2},
but now with a higher signal-to-noise ratio.
}

\clearpage

%%%%%%%%%%%%%%%%%%%%%%%%%%%%%%%%%%%%%%%%%%%%%%%%%%%%%%%%%%%%%%%%%

%%% HACK: use to create figures with no page numbers ...
\pagestyle{empty}

%%%%%%%%%%%%%%%%% Figure 1 %%%%%%%%%%%%%%%%%

\newpage
\clearpage

%%%%%%%%%%%%%%%%% Figure 2 %%%%%%%%%%%%%%%%%

\newpage
\clearpage

%%%%%%%%%%%%%%%%% Figure 3 %%%%%%%%%%%%%%%%%

\newpage
\clearpage

%%%%%%%%%%%%%%%%% Figure 4 %%%%%%%%%%%%%%%%%

\newpage
\clearpage

%%%%%%%%%%%%%%%%% Figure 5 %%%%%%%%%%%%%%%%%

\newpage
\clearpage

%%%%%%%%%%%%%%%%% Figure 6 %%%%%%%%%%%%%%%%%

\newpage
\clearpage

%%%%%%%%%%%%%%%%% Figure 7 %%%%%%%%%%%%%%%%%

\newpage
\clearpage

%%%%%%%%%%%%%%%%% Figure 8 %%%%%%%%%%%%%%%%%

\newpage
\clearpage

\end{document}